\documentclass[twocolumn,superscriptaddress,showpacs,preprintnumbers,amsmath,amssymb,prl]{revtex4}
\usepackage{graphicx}% Include figure files
\usepackage{dcolumn}% Align table columns on decimal point
\usepackage{bm}% bold math

\begin{document}

%\preprint{APS/123-QED}

\title{Damping of Bogoliubov Excitations in Optical Lattices}% Force line breaks with \\

\author{Shunji Tsuchiya}
% \email{tsuchiya@physics.utoronto.ca}
\affiliation{Department of Physics, University of Toronto, Toronto, Ontario, Canada M5S 1A7}
\affiliation{Department of Physics, Waseda University, 3-4-1 Okubo, Tokyo 169-8555, Japan.}
%\altaffiliation[Also at ]{Department of Physics, Waseda University, 3-4-1 Okubo, Tokyo 169-8555, Japan.}%Lines break automatically or can be forced with \\
\author{Allan Griffin}%
%\email{griffin@physics.utoronto.ca}
\affiliation{Department of Physics, University of Toronto, Toronto, Ontario, Canada M5S 1A7}

\date{\today}% It is always \today, today,

\begin{abstract}
Extending recent work to finite temperatures, we calculate the Landau damping of a Bogoliubov excitation in an optical lattice, due to coupling to a thermal cloud of such excitations. 
For simplicity, we consider a 1D Bose-Hubbard model and restrict ourselves to the first energy band.
For energy conservation to be satisfied, the excitations 
in the collision processes must exhibit ``anomalous dispersion'', 
analogous to phonons in superfluid $^4\rm{He}$.  
This leads to the disappearance of all damping processes 
when $U n^{\rm c 0}\ge 6t$, where $U$ is the on-site interaction, 
$t$ is the hopping matrix element and $n^{\rm c 0}(T)$ is 
the number of condensate atoms at a lattice site. 
This phenomenon also occurs in 2D and 3D optical lattices.
The disappearance of Beliaev damping above a threshold wavevector is noted.
\end{abstract}

\maketitle

%\section{\label{sec:level1}Introduction}

Recently, several experimental papers have reported results on the damping of collective modes in Bose condensates 
in a one-dimensional (1D) periodic optical lattice potential \cite{Inguscio2}.
Most theoretical studies \cite{Smerzi2,Wu,Machholm,Jaksch,ReyBurnett,Stoof} have concentrated on the case of a pure condensate at $T=0$ 
and have ignored the effect of a thermal cloud, which can cause damping of condensate excitations at finite $T$.
We discuss a Bose-Hubbard tight-binding model using the Gross-Pitaevskii (GP) approximation,
but generalized to include a static thermal cloud of non-condensate atoms.
For illustration, we consider a 1D model but emphasize the same features appear in 2D and 3D optical lattices \cite{TsuchiyaGriffin}.
We limit our analysis to relatively strong optical lattices such that the lowest energy band is fairly narrow.
We calculate the temperature dependence of 
the number of condensate atoms $n^{\rm c 0}(T)$ in each lattice well, 
as well as the Landau damping of 
condensate modes due to coupling to thermal excitations.
We find that for damping processes to occur, 
the dispersion relation $E_q$ of the Bogoliubov Bloch excitations 
must initially bend upward as the quasi-momentum $q$ increases. 
This ``anomalous dispersion'' is also the source of 3-phonon damping 
in superfluid $^4\rm{He}$ \cite{anomalous1,anomalous2}.
This condition leads to the disappearance of all damping in an 1D optical lattice when the normalized interaction strength $\alpha\equiv Un^{\rm c 0}/t >6$ 
(where $U$ is the on-site interaction and $t$ is the hopping matrix element).
More generally, all damping processes involving three excitations
vanish in a $D$-dimensional simple cubic tight-binding optical lattice when
$\alpha>6D$ \cite{TsuchiyaGriffin}. 
This ability to control the excitation damping in optical lattices may be very important in applications.

Since our major interest is in the thermal gas of excitations 
in an optical lattice, 
it is important to keep in mind the distinction 
between the Bloch excitations associated with linearized fluctuations 
of an equilibrium Bose-condensate and the stationary states of 
the time-independent GP equation for the Bose order parameter.
In a continuum model, the latter states \cite{Wu,Machholm} can be described by the eigenfunctions
$\Phi^0_k(x)=e^{\mathrm{i}kx}u_k(x)$,
where the Bloch condensate function satisfies the usual 
periodicity condition $u_k(x)=u_k(x+ld)$, where $d$ is the optical
lattice spacing and $l$ is an integer.
Physically, $\Phi_k^0(x)$ corresponds to a solution of 
the static GP equation with a superfluid flow in the periodic potential, 
with the condensate quasi-momentum being given by $\hbar k$.
The fluctuations of these states are also described by a quasi-momentum $q$ 
in the first Brillouin zone (BZ) and will be referred to as Bogoliubov Bloch
excitations of the optical lattice.
The thermal cloud at finite temperatures 
is an incoherent gas of these Bogoliubov excitations in the first band.
In our tight-binding model, the Bloch condensate function is 
$\Phi_k^0(l)=e^{\mathrm{i}kld}\sqrt{n^{\mathrm c 0}}$
and we only consider a Bose condensate 
in the $k=0$ Bloch state, {\it i.e.}, 
$\Phi_{k=0}^0(l)=\sqrt{n^{\mathrm{c}0}}$.

The 1D optical lattice model (along the $z$ axis) has been discussed in many recent papers at $T=0$ \cite{Smerzi2,Jaksch,ReyBurnett,Stoof}.
The optical lattice is described by 
$V_\mathrm{op}(z)=s E_R\cos^2(kz)$, where $E_R=\frac{\hbar^2k^2}{2m}$ is 
the photon recoil energy. 
We assume that the band gap energy between the first and the second excitation band is large compared to the temperature ($2k_\mathrm{B}T/E_R\ll s$), and thus only the first band 
is thermally occupied.
The radial trapping frequency is assumed to be very large, so that 
motion in this direction is effectively frozen.
The depth of the optical lattice wells is assumed large enough to make the atomic wave functions 
well localized on the individual lattice sites, described by a tight-binding 
approximation. This Hamiltonian is the Bose-Hubbard model, 
%%%%%%%%%%%%%%%%%%%%%%% Bose-Hubbard %%%%%%%%%%%%%%%%%%%%%%%%%%%%%%%%
\begin{eqnarray}
H=-t\sum_{l}(a_{l+1}^\dagger a_{l}^{}+a_{l}^\dagger a_{l+1}^{})
+\frac{1}{2}U\sum_{l} a_{l}^\dagger a_{l}^\dagger
a_{l}^{} a_{l}^{},\label{H}
\end{eqnarray}
%%%%%%%%%%%%%%%%%%%%%%%%%%%%%%%%%%%%%%%%%%%%%%%%%%%%%%%%%%%%%%%%%%%%
where $a_l$ and $a_l^\dagger$ are creation and destruction operators 
of atoms in the radial ground state on the $l$-th lattice site. 
The hopping matrix element between adjacent sites is 
$t=-\int dz w_l^\ast(z) \left( -\frac{\hbar^2}{2m}\frac{d^2}{dz^2}
+V_\mathrm{op}(z)\right)w_{l+1}(z)$, where $w_l(z)$ is a 
function localized on the $l$-th lattice site. 
The on-site inter-atom interaction is $U=g \int d\mathbf{r}_\bot \left|\phi_0(\mathbf{r}_\bot)\right|^4\int d{z}|w_l(z)|^4$,
 where $g=4\pi\hbar^2a/m$ and $a$ is the $s$-wave scattering length. 
$\phi_0(\mathbf{r}_\bot)$ is the wave function 
in the radial direction. 

Expanding around the minima of the optical lattice potential wells in a harmonic approximation, the well trap frequency is 
$\omega_s\equiv s^{1/2}\frac{\hbar k^2}{m}$, 
which is assumed to be much larger than the harmonic trapping frequency 
along the $z$-axis.
Approximating the localized atomic wave function as a gaussian at the potential minima of $l$-th site, 
$w_l(z)=\left(\frac{m\omega_s}{\pi\hbar}\right)^{1/4}\exp\left(-\frac{m\omega_s}{2\hbar}(z-z_l)^2\right)$, one obtains
%%%%%%%%%%%%%%%%%%%%%%% hopping matrix element %%%%%%%%%%%%%%%%%%%%%%%%%%
\begin{eqnarray}
\frac{t}{E_\mathrm{R}}\sim\left[\big(\frac{\pi^2 s}{4}-\frac{s^{1/2}}{2}\big)-\frac{1}{2}s(1+e^{-s^{-1/2}})\right]e^{-\pi^2 s^{1/2}/4}.\label{t}
\end{eqnarray}
%%%%%%%%%%%%%%%%%%%%%%%%%%%%%%%%%%%%%%%%%%%%%%%%%%%%%%%%%%%%%%%%%%%%%%%%%
Taking the wave function in the radial direction to be the ground state wave 
function of a harmonic oscillator, one obtains ($d=\pi/k$ is the lattice period)
%%%%%%%%%%%%%%%%%%%%%%%%% on-site interaction %%%%%%%%%%%%%%%%%%%%%%%%%%%
\begin{eqnarray}
\frac{U}{E_\mathrm{R}}\sim&&\frac{g}{(2\pi)^{3/2}a_\bot^2 a_s}
=\frac{2^{3/2}a d}{\pi^{3/2}a_\bot^2}s^{1/4},\label{U}
\end{eqnarray}
%%%%%%%%%%%%%%%%%%%%%%%%%%%%%%%%%%%%%%%%%%%%%%%%%%%%%%%%%%%%%%%%%%%%%%%%%
where $a_s=\sqrt{\frac{\hbar}{m\omega_s}}$, $a_\bot=\sqrt{\frac{\hbar}{m\omega_\bot}}$.
%%%%%%%%%%%%%%%%%%%%%%%%%%%%%%%%%%%%%%%%%%%%%%%%%%%%%%%%
%\begin{figure}[keepaspectration]
%\includegraphics[height=4.5cm]{fig1.eps}
%\includegraphics[width=\linewidth]{fig1.eps}
%\caption{Plot of the ratio $\alpha=Un^{\mathrm c 0}(T)/t$,
%as a function of temperature, for several value of the strength of the optical lattice potential $s$.
%The number of condensate atoms at a lattice site
%$n^{\mathrm c 0}(T)$ is given in Fig. 3.
%The dashed line is $\alpha=6$.} 
%\end{figure}
%%%%%%%%%%%%%%%%%%%%%%%%%%%%%%%%%%%%%%%%%%%%%%%%%%%%%%%%
%Using these results, we plot the dimensionless interaction $\alpha\equiv Un^{\mathrm c 0}/t$ 
%as a function of temperature in Fig. 1.
%The ratio $\alpha$ and the results in Fig. 1 will play a crucial role in our subsequent analysis.

We restrict ourselves to the superfluid solutions of
Eq. (\ref{H}), when the phase coherence between wells is well defined in the optical lattice.
The superfluid-Mott transition \cite{Jaksch,Stoof,Greiner} only arises when the interaction energy is much larger than we consider here.
Due to Bose condensation, the destruction operator of an atom at site $l$ 
can be written as 
$a_l=\Phi_l+\tilde{\psi}_l$, 
where $\Phi_l=\langle a_l\rangle$ is the condensate wave function and 
$\tilde{\psi}_l$ is the non-condensate field operator.
This leads to the generalized discrete Gross-Pitaevskii equation \cite{Hutchinson} 
(we set $\hbar=1$ from now on),
%%%%%% Discretized Gross-Pitaevskii equation %%%%%%%%%%%%%%%%%%%
\begin{eqnarray}
\mathrm{i}\frac{\partial}{\partial t}\Phi_l=-t(\Phi_{l+1}+\Phi_{l-1})
+U(n_l^\mathrm{c}+2\tilde{n}_l)\Phi_l, \label{GP}
\end{eqnarray}
%%%%%%%%%%%%%%%%%%%%%%%%%%%%%%%%%%%%%%%%%%%%%%%%%%%%%%%%%%%%%%%%
where $n_l^\mathrm{c}(t)$ is the number of condensate atoms on the $l$-th 
lattice site.
This includes the time-dependent Hartree-Fock mean field $2U\tilde{n}_l(t)$ arising from the non-condensate atoms on the $l$-th site.
Eq. (\ref{GP}) reduces to the usual discrete tight binding 
GP equation \cite{Smerzi2,ReyBurnett} if all the atoms are 
assumed to be in the condensate ( {\it i.e.}, $\tilde{n}_l=0$).
Introducing phase and amplitude variables, $\Phi_{l}=\sqrt{n_l^\mathrm{c}}e^{\mathrm{i}\theta_{l}}$, Eq. (\ref{GP}) reduces to
%%%%%%%%%%%% continuity equation 1 %%%%%%%%%%%%%%%%%%%%%%%%%%
\begin{eqnarray}
\frac{\partial n_l^\mathrm{c}}{\partial t}&=&-2t\sqrt{n_l^\mathrm{c} n_{l+1}^\mathrm{c}}\sin(\theta_{l+1}-\theta_l)\nonumber\\
& & {} +2t\sqrt{n_l^\mathrm{c} n_{l-1}^\mathrm{c}}\sin(\theta_{l}-\theta_{l-1}), \label{continuity}
\end{eqnarray}
%%%%%%%%%%%%%%%%%%%%%%%%%%%%%%%%%%%%%%%%%%%%%%%%%%%%%%%%%%%%%%%%
%%%%%% Josephson equation 1  %%%%%%%%%%%%%%%%%%%%%%%%%%%%%%%%%%%%
\begin{eqnarray}
\frac{\partial \theta_l}{\partial t}
&=&t\left(\sqrt{\frac{n_{l+1}^\mathrm{c}}{n_l^\mathrm{c}}}\cos(\theta_{l+1}-\theta_{l})
+\sqrt{\frac{n_{l-1}^\mathrm{c}}{n_l^\mathrm{c}}}\cos(\theta_{l}-\theta_{l-1})\right)\nonumber\\
& & {} -(Un_l^\mathrm{c}+2U\tilde{n}_l)\equiv -\varepsilon_l^\mathrm{c}.
\label{Josephson}
\end{eqnarray}
%%%%%%%%%%%%%%%%%%%%%%%%%%%%%%%%%%%%%%%%%%%%%%%%%%%%%%%%%%%%%
In Eq. (\ref{Josephson}), $\varepsilon_l^\mathrm{c}$ is an energy of 
a condensate atom, which reduces 
to the chemical potential of the condensate 
$\mu_{\mathrm c 0}$ in static thermal equilibrium.
The equilibrium solution ($\theta_l$ and $n_l^{\mathrm c}$ are independent of $l$) is 
$\theta_l(t)=-\mu_{\mathrm c 0}t$.
%From Eq. (\ref{continuity}), the Josephson superfluid current 
%between $l$-th and $(l-1)$-th lattice sites is
%$J_l=2t\sqrt{n_l^\mathrm{c}n_{l-1}^\mathrm{c}}\sin(\theta_l-\theta_{l-1})$. It vanishes in equilibrium.

The $T=0$ Bogoliubov excitation spectrum \cite{Javanainen,Stoof,ReyBurnett} for a uniform optical lattice is easily obtained 
by ignoring the non-condensate atom term ($\tilde{n}_l=0$) 
and considering small fluctuations from equilibrium, $n_l^\mathrm{c}=n^{\mathrm{c}0}+\delta n_l^\mathrm{c}$.
%%%%%%%%%%%% continuity equation 2 %%%%%%%%%%%%%%%%%%%%%
%\begin{eqnarray}
%\frac{\partial\delta n_l^\mathrm{c}}{\partial t}
%+2t n^{\mathrm{c}0}\left(\delta \theta_{l+1}-2\delta \theta_{l+1}+\delta \theta_{l-1}\right)=0, \label{continuity2}
%\end{eqnarray}
%%%%%%%%%%%%%%%%%%%%%%%%%%%%%%%%%%%%%%%%%%%%%%%%%%%%%%%%
%%%%%%%%%%%% Josephson equation 2 %%%%%%%%%%%%%%%%%%%%%%
%\begin{eqnarray}
%\frac{\partial\delta \theta_l}{\partial t}=\frac{t}{2n^{\mathrm{c}0}}\left(\delta n_{l+1}^\mathrm{c}-2\delta n_l^\mathrm{c}+\delta n_{l-1}^\mathrm{c}\right)-U\left(\delta n_l^{\mathrm{c}}+2\delta\tilde{n}_l\right).\label{Josephson2}
%\end{eqnarray}
%%%%%%%%%%%%%%%%%%%%%%%%%%%%%%%%%%%%%%%%%%%%%%%%%%%%%%%%
The normal mode solutions of Eqs. (\ref{continuity}) and (\ref{Josephson}) 
are  
%%%%%%%%%%%%%  plane wave solution  %%%%%%%%%%%%%%%%%%%%%%%
\begin{eqnarray}
\delta \theta_l=\delta\theta(q)e^{{\mathrm i} [qld-E_qt]},\ \ 
\delta n_l^\mathrm{c}=\delta n^{\mathrm c}(q)e^{{\mathrm i} [qld-E_qt]},
\end{eqnarray} 
%%%%%%%%%%%%%%%%%%%%%%%%%%%%%%%%%%%%%%%%%%%%%%%%%%%%%%%%%%%
with the Bogoliubov Bloch excitation energy,
%%%%%%%%%%%%  Bogoliubov spectrum   %%%%%%%%%%%%%%%%%%%%%%
\begin{eqnarray}
E_q=\sqrt{\epsilon_q^0\left(\epsilon_q^0+2Un^{\mathrm{c}0}\right)}\label{Bogoliubov}.
\end{eqnarray}
%%%%%%%%%%%%%%%%%%%%%%%%%%%%%%%%%%%%%%%%%%%%%%%%%%%%%%%%%
Here $\epsilon_q^0\equiv 4t\sin^2\frac{qd}{2}$ is the kinetic energy associated with tunneling.
This $T=0$ excitation spectrum is shown in Fig. 1 for two values of the interaction ratio $\alpha\equiv Un^\mathrm{c 0}/t$. 
For small $q$, the spectrum is phonon-like $E_q\simeq cq$, with the phonon speed $c=\sqrt{2td^2Un^{\mathrm{c}0}}=\sqrt{\frac{Un^{\mathrm{c}0}}{m^\ast}}$, where $m^\ast\equiv\frac{1}{2td^2}$ is an effective mass of atoms in the optical lattice.
%%%%%%%%%%%%%%%%%%%%%%%%%%%%%%%%%%%%%%%%%%%%%%%%%%%%%%%%
\begin{figure}[keepaspectratio]
\includegraphics[height=4.5cm]{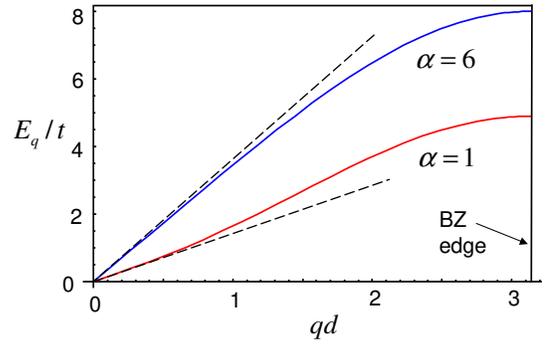}
\caption{The Bogoliubov excitation energy $E_q$ (normalized to $t$) in the first Brillouin zone, for $\alpha=1$ and $6$, as a function of the quasi-momentum $q$. The dashed lines give $E_q=cq$.} 
\vspace*{-0.3cm}
\end{figure}
%%%%%%%%%%%%%%%%%%%%%%%%%%%%%%%%%%%%%%%%%%%%%%%%%%%%%%%%
We call attention to an important feature of Eq. (\ref{Bogoliubov}). 
For $\alpha \le 6$, $E_q$ bends up 
before bending over, as $q$ approaches the BZ boundary.
This ``anomalous dispersion'' also occurs in superfluid $^4\mathrm{He}$ \cite{anomalous1,anomalous2}.
For $\alpha>6$, the spectrum simply bends over as $q$ increases. 

As we noted earlier, the energy of a condensate atom $\varepsilon_l^\mathrm{c}$ on site $l$ is given by the r.h.s of Eq. (\ref{Josephson}).
In static thermal equilibrium, when all sites are identical, 
the condensate chemical potential is given by 
$\mu_{\mathrm{c}0}=\varepsilon^{\mathrm c 0}=
-2t+U[n^{\mathrm c 0}(T)+2\tilde n^0(T)]$.
We calculate $n^{\mathrm c 0}(T)$ self-consistently by calculating 
the number of non-condensate atoms $\tilde{n}^\mathrm{0}(T)$ at a site using the Bogoliubov excitations given by the static Popov approximation.
That is, we ignore the dynamics of the non-condensate [{\it i.e.}, we set $\delta\tilde{n}_l(t)=0$ in Eq. (\ref{Josephson})].
This gives the Bogoliubov-Popov excitation spectrum $E_q$ which is 
identical to Eq. (\ref{Bogoliubov}), except that now $n^{\mathrm c 0}(T)$ is 
the temperature-dependent number of condensate atoms at any lattice site.

Expressing $\tilde{n}^0(T)$ in terms of these Bogoliubov-Popov excitations, 
we have \cite{ReyBurnett,Stoof,Hutchinson}
%%%%%%%%%% Bogoliubov approximation %%%%%%%%%%%%%%%%%%%%
\begin{eqnarray}
n=n^{\mathrm{c}0}+\frac{1}{I}\sum_{|q|\ge q_\mathrm{c}}\left[\left(u_q^2+v_q^2\right)f^{0}(E_q)+v_q^2\right],\label{selfconsist}
\end{eqnarray}
%%%%%%%%%%%%%%%%%%%%%%%%%%%%%%%%%%%%%%%%%%%%%%%%%%%%%%%%
with the standard Bogoliubov transformation functions
$u_q^2=1/2(\tilde{E}_q/E_q+1)$, $v_q^2=1/2(\tilde{E}_q/E_q-1)$,
and $f^{0}(E_q)=[\exp(\beta E_q)-1]^{-1}$.
The Hartree-Fock (HF) excitation spectrum is
$\tilde{E}_q\equiv[\epsilon_q^0+2U(n^{\mathrm c 0}+\tilde{n}^0)]-\tilde{\mu}_0=4t\sin^2(\frac{qd}{2})+Un^{\mathrm c 0}$. Apart from the limiting case of $\alpha \ll 1$,
we note that $E_q$ is quite different from the HF spectrum $\tilde{E}_q$.
Strictly speaking, in our 1D model, there are no solutions of Eq. (\ref{selfconsist}) for an infinite system because 
the summation diverges due the contribution from small momentum, 
in accordance with the Mermin-Wagner-Hohenberg 
theorem.
However, we consider finite systems which introduces a lower momentum cutoff $q_\mathrm{c}\equiv 2\pi/Id$. (As a typical value \cite{Inguscio2}, we take the total number of lattice sites $I=250$.)  
In this case, Eq. (\ref{selfconsist}) has a solution and the condensate 
can exist. We are only considering a 1D optical lattice
for illustration and emphasize that our main conclusions
also hold for 2D and 3D optical lattices  \cite{TsuchiyaGriffin}. 

The condensate number $n^{\mathrm{c}0}(T)$ is found by solving Eq. (\ref{selfconsist}) 
self-consistently for a fixed value of $n$. 
In Fig. 2, we plot $n^{\mathrm c 0}(T)/n$ as a function of $T$. 
Using these results, we plot the dimensionless interaction $\alpha\equiv Un^{\mathrm c 0}(T)/t$ as a function of temperature in Fig. 3.
The ratio $\alpha$ and the results in Fig. 3 will play a crucial role in our subsequent analysis. Since we limit our discussion to the first energy band of the optical lattice, our results only apply when $s\gg 2k_\mathrm{B}T/E_R$.
Higher bands would be thermally populated if we considered lower values of $s$.
%At finite $T$, the dimensionless interaction is defined as $\alpha\equiv Un^{\mathrm c 0}(T)/t$. 
%%%%%%%%%%%%%%%%%%%%%%%%%%%%%%%%%%%%%%%%%%%%%%%%%%%%%%%%%
\begin{figure}[keepaspectratio]
\includegraphics[height=4.5cm]{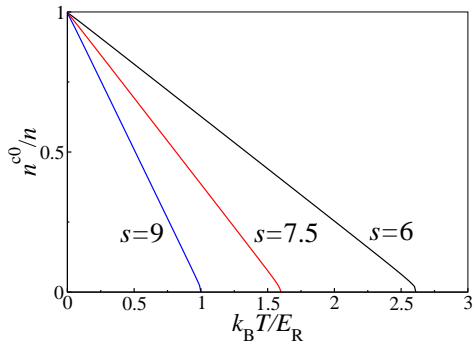}
\caption{The site condensate fraction $n^{\mathrm c 0}/n$ as a function of temperature, where $s$ is the strength of the optical lattice potential. Following Ref. \cite{Inguscio2}, we take $n=1200$.}
\vspace*{-0.5cm}
\end{figure}
%%%%%%%%%%%%%%%%%%%%%%%%%%%%%%%%%%%%%%%%%%%%%%%%%%%%%%%%
%%%%%%%%%%%%%%%%%%%%%%%%%%%%%%%%%%%%%%%%%%%%%%%%%%%%%%%%
\begin{figure}[keepaspectration]
\includegraphics[height=4.5cm]{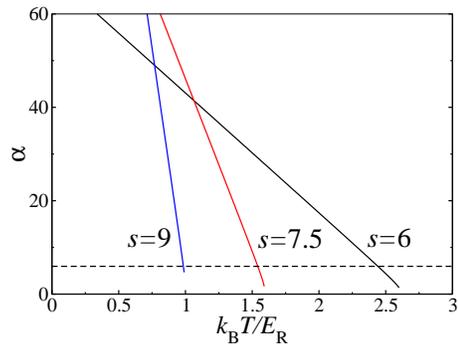}
\caption{Plot of the interaction parameter $\alpha=Un^{\mathrm c 0}(T)/t$,
as a function of temperature, for several values of $s$.
The number of condensate atoms at a lattice site
$n^{\mathrm c 0}(T)$ is given in Fig. 2.
The dashed line is $\alpha=6$.} 
\vspace*{-0.3cm}
\end{figure}
%%%%%%%%%%%%%%%%%%%%%%%%%%%%%%%%%%%%%%%%%%%%%%%%%%%%%%%%
%
%It is easy to verify that $E_q$ in Eq. (\ref{Bogoliubov}) reduces to 
%the HF energy $\tilde{E}_q$ in the limit $t \gg U n^{\mathrm c 0}$, {\it i.e.}, when $\alpha\ll 1$.
%This HF limit corresponds to setting the Bogoliubov amplitudes $u_q^2=1$, $v_q^2=0$.
%In dealing with Bose gases trapped in harmonic potential wells,
%one can always \cite{ZNG} use the high energy HF approximation for the excitations describing 
%the thermal cloud as long as the kinetic energy of the atoms ($\sim k_\mathrm{B} T$) 
%is much larger than the interaction energy ($U n^{\mathrm c 0}$).
%In contrast, apart from the limiting case of $\alpha\ll 1$,
%we must always use the full Bogoliubov spectrum $E_q$ to describe the thermal cloud in the first excitation band of an optical lattice.

We next turn to the calculation of Landau damping.
%%%%%%%%%%%%%%%%%  Pitaevskii Stringari's method  %%%%%%%%%%%%%%%%%%%%%%%%%%%
A collective mode in a condensate coupled to a thermal cloud 
of excitations in thermal equilibrium has
a complex frequency, $\omega\equiv E_q-\mathrm{i} \Gamma_q$, with the Landau damping given by \cite{Pethick&Smith,PitaevskiiStringari}
%%%%%%%%%%%%%%%%%%%%%%%% Landau damping %%%%%%%%%%%%%%%%%%%%%%%%
\begin{eqnarray}
\Gamma_q&=&\pi\sum_{|p|\ge q_\mathrm{c}} |M_{q,p;q+p}|^2[f^0\left(E_{p}\right)-f^0\left(E_{q+p}\right)]\nonumber\\
& & {} \times\delta\left(E_q+E_{p}-E_{q+p}\right).\label{Landaudamp*}
\end{eqnarray}
%%%%%%%%%%%%%%%%%%%%%%%%%%%%%%%%%%%%%%%%%%%%%%%%%%%%%%%%%%%%%%%%%
Here $E_p$ is the Bloch Bogoliubov excitation energy given in Eq. (\ref{Bogoliubov}) for a uniform condensate in an optical lattice, 
and the momentum sum is over the first Brillouin zone.
%We only consider damping from the lowest excitation energy band.
The expression in Eq. (\ref{Landaudamp*}) describes a condensate excitation of (quasi) momentum $q$ being absorbed by an excitation $p$ of the optical lattice thermal gas, leading to a thermal excitation with momentum $q+p$.

%%%%%%%%%%%%%% Peierls's graphycal method %%%%%%%%%%%%%%%%%%%%%%%%%%%%%
The energy conservation condition $E_q+E_p=E_{q+p}$ needs to be satisfied in Eq. (\ref{Landaudamp*}). 
This is illustrated in Fig. 4 \cite{Peierls}.
Clearly, the intersection at $(q+p, E_{q+p})$ requires that the dispersion relation $E_q$ first bends up 
as $q$ increases, before bending over. 
This anomalous dispersion is also a feature of phonons in superfluid $^4\rm{He}$ \cite{anomalous1,anomalous2}.
%%%%%%%%%%%%%%%%%%%%%%%%%%%%%%%%%%%%%%%%%%%%%%%%%%%%%%%%%%%%%%%%%%%%%%%
%%%%%%%%%%%%%%%% explanation of Fig. 5 %%%%%%%%%%%%%%%%%%%%%%%%%%%%%%%%%
The values of $(q,p)$ satisfying $E_q+E_p=E_{q+p}$ is shown in Fig. 5. 
For a given $q$, we see that as $\alpha \to 6$, the value of $p$ decreases to zero. 
There is no solution for $\alpha>6$, indicating the disappearance of Landau damping.
%%%%%%%%%%%%%%%%%%%%%%%%%%%%%%%%%%%%%%%%%%%%%%%%%%%%%%%%%%%%%%%%%%%%%%%%%
%%%%%%%%%%%%%%%%%%%%%%%%%%%%%%%%%%%%%%%%%%%%%%%%%%%%%%%%%%%%%%%%%%%%%%%  
%The matrix element $M_{q,p_1\ ;\ p_2}$ describing this 3-excitation process 
%is calculated by extracting the terms proportional to $\alpha_{p_2}^\dagger\alpha_{p_1}\alpha_{q}$ in the interaction term of Eq. (\ref{H}) \cite{PitaevskiiStringari}, where $\alpha_p$ and $\alpha_p^{\dagger}$ are the creation 
%and destruction operators of elementary excitations.

The matrix element in Eq. (\ref{Landaudamp*}) is given by \cite{PitaevskiiStringari,Pethick&Smith}
%%%%%%%%%%%%%% matrix element by Pitaevskii and Stringari %%%%%%%%%%%%%%%
\begin{eqnarray}
M_{q,p_1; p_2}=2U\sqrt{\frac{n^{\mathrm c 0}}{I}}\sum_{G_n}
\big[(u_{p_1}u_{p_2}+v_{p_1}v_{p_2}-v_{p_1}u_{p_2})u_q\nonumber\\
-(u_{p_1}u_{p_2}+v_{p_1}v_{p_2}-u_{p_1}v_{p_2})v_q\big]
\delta_{p+p_1,\ p_2+G_n},\hspace{0.5cm} \label{matrixelement1*}
\end{eqnarray}
%%%%%%%%%%%%%%%%%%%%%%%%%%%%%%%%%%%%%%%%%%%%%%%%%%%%%%%%%%%%%%%%%%%%%%%%%
where $G_n=\frac{2\pi n}{d}$ is a reciprocal lattice vector 
($n$ is an integer). In Fig. 5, the curves of the solution of the energy conservation condition never go across the dashed line $q+p=\pi/d$. 
Thus Umklapp scattering processes ($G_n\ne 0$) do not contribute to Landau damping \cite{Peierls}.

When the excitation $q$ has a wavelength much larger than
the thermal excitation $p$ ({\it i.e.} $q\ll p$, $\pi/d$),
the energy conservation condition $E_q+E_{p}=E_{q+p}$ in Eq. (\ref{Landaudamp*})
means that for $q\ll p$, the Landau damping of the excitation $E_q=cq$ comes from absorbing a thermal excitation $E_{p}$ with a group velocity $\frac{\partial E_{p}}{\partial p}$ equal to $c$ \cite{Pitaevskii&Levinson}. 
The unique value of $p_0$ such that 
$\frac{\partial E_{p_0}}{\partial p_0}=c$ is found to be given by the condition $4\sin^2(\frac{1}{2}p_0d)=-(\alpha-2)+\sqrt{2(\alpha+2)}$. This is only valid for $\alpha$ smaller than 6, such that $p_0\gg q$.

For $q\ll p$, the matrix element reduces to 
(the analogue of Eq. (10.76) in Ref. \cite{Pethick&Smith} )
%%%%%%%%%%%%%%%%%%%%%%%%%%%%%%%%%%%%%%%%%%%%%%%%%%%%%%%%%%%%%%%%%%%%%%%
\begin{eqnarray}
M_{q,p;q+p}=U\left(\frac{\epsilon_p^0}{E_p}+\frac{E_p}{\tilde{E}_p}\right)\left(\frac{q}{2m^\ast c}\right)^{1/2}\frac{N_{\mathrm c 0}^{1/2}}{I},\label{matrixelement2*}
\end{eqnarray}
%%%%%%%%%%%%%%%%%%%%%%%%%%%%%%%%%%%%%%%%%%%%%%%%%%%%%%%%%%%%%%%%%%%%%%%
where $N_{\mathrm c 0}\equiv n^{\mathrm c 0}I$. 
In deriving Eq. (\ref{matrixelement2*}) from Eq. (\ref{matrixelement1*}), 
we have used the energy conservation factor in Eq. (\ref{Landaudamp*}).
%%%%%%%%%%%%%%%%%%%%%%%%%%%%%%%%%%%%%%%%%%%%%%%%%%%%%%%%%%%%%%%%%%%%%%%%
Using the delta function for energy conservation 
to integrate over $p$, the Landau damping is given by 
%%%%%%%%%%%%%%%%%%%%%%%%%%%%%%%%%%%%%%%%%%%%%%%%%%%%%%%%%%%%%%%%%%%%%%%
\begin{eqnarray}
\Gamma_q=\frac{\alpha}{8(\alpha+2)}\left(\frac{E_{p_0}}{\epsilon^0_{p_0}} \right)^3\left(\frac{\epsilon^0_{p_0}}{E_{p_0}}+\frac{E_{p_0}}{\tilde{E}_{p_0}} \right)^2\frac{\beta t Uqd}{\sinh^2\frac{\beta E_{p_0}}{2}}.\hspace{0.3cm}\label{Landaudamping}
\end{eqnarray}
%%%%%%%%%%%%%%%%%%%%%%%%%%%%%%%%%%%%%%%%%%%%%%%%%%%%%%%%%%%%%%%%%%%%%%%

%%%%%%%%%%%%%%%%%%%% Peierls graphycal method %%%%%%%%%%%%%%%%%%%%%%%%%
\begin{figure}[keepaspectratio]
\includegraphics[height=4cm]{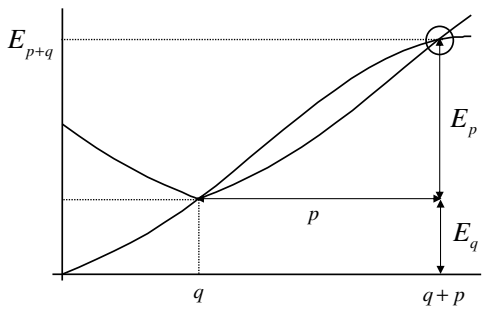}
\caption{The Bogoliubov excitation energy $E_q$ for $\alpha<6$. The intersection of the two dispersion curves at $p+q$ shows that the energy conservation condition is satisfied.}
\end{figure}
%%%%%%%%%%%%%%%%%%%%%%%%%%%%%%%%%%%%%%%%%%%%%%%%%%%%%%%%%%%%%%%%%%%%%%%
%%%%%%%%%%%%%%%%%%%% energy conservation condition %%%%%%%%%%%%%%%%%%%
\begin{figure}[keepaspectratio]
\vspace*{0.3cm}
\includegraphics[height=4cm]{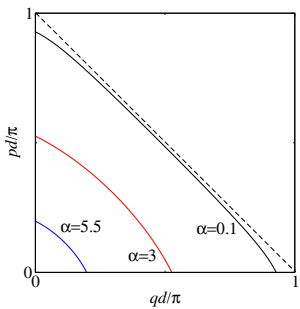}
\caption{The values $(q,p)$ satisfying the energy conservation condition $E_q+E_p=E_{q+p}$, for several values of $\alpha$.}
\vspace{-0.2cm}
\end{figure}
%%%%%%%%%%%%%%%%%%%%%%%%%%%%%%%%%%%%%%%%%%%%%%%%%%%%%%%%%%%%%%%%%%%%%% 

%%%%%%%%%%%%%%%%%%%%%%%%%%%%%%%%%%%%%%%%%%%%%%%%%%%%%%%%%%%%%%%%%%%%%%%
%\begin{figure}
%\includegraphics[height=5cm,width=7.1cm]{fig6.eps}
%\includegraphics[width=\linewidth]{fig6.eps}
%\caption{The Landau damping $\Gamma_q$ of phonon excitation $E_q=cq$, 
%as a function of the optical lattice well depth $s$.
%Results for several temperatures are shown.}
%\end{figure}
%%%%%%%%%%%%%%%%%%%%%%%%%%%%%%%%%%%%%%%%%%%%%%%%%%%%%%%%%%%%%%%%%%%%%%%
One finds there is no damping when $\alpha>6$ (see Fig. 3 for $\alpha$ as a function of $s$ and $T$). $\Gamma_q$ diverges at $\alpha=6$, but this is due to the 1D nature of our system and does not occur in 2D lattices.
Fig. 3 shows that, for $n=1200$ \cite{Inguscio2}, $\alpha$ is smaller than 6 only when $T$ is very close to $T_\mathrm{c}$, for $s$ in the range $6\sim9$.
For a smaller number of atoms on a lattice site, the temperature range where $\alpha$ is less than 6 becomes larger.
%Fig. 3 shows that for $k_\mathrm B T \agt 10E_\mathrm R$, 
%$\alpha$ is always smaller than 6. 
%The interesting feature is that, for $k_{\mathrm B} T \alt 10 E_{\mathrm R}$, 
%there is always an intermediate range of values of the optical lattice depth $s$ such that $\alpha>6$, where Landau damping vanishes. 

%We have extended \cite{TsuchiyaGriffin} the present calculation to the damping of excitations of an equilibrium condensate with $k\neq 0$, 
%as described by Eq. (\ref{Bloch_condensate}).
%All previous work \cite{Smerzi2,Wu,Machholm} on this problem only considered dynamical instabilities in the absence of any real dissipation mechanisms
%such as treated in the present work.

For our 1D model to apply, one would need a much tighter magnetic trap (in the radial direction) than used in Ref.\cite{Inguscio2}.
A 3D optical lattice (without any magnetic trap) 
might allow the clearest check of our predictions of the disappearance of damping.

One also has damping from collisions which transfer atoms between the condensate and thermal cloud \cite{WG1} and Beliaev damping (present even at $T=0$) involving the spontaneous decay of an excitation into two excitations.
All such processes \cite{TsuchiyaGriffin} also involve an energy conservation factor of the kind $\delta(E_1-E_2-E_3)$, which can only be satisfied 
if the excitations exhibit anomalous dispersion, $i.e.$, $\alpha\le 6D$, where $D$ is the dimension of the optical lattice. 
Finally, for a fixed value of $\alpha<6D$, there is a finite threshold momentum $q^\ast$ such that the decay (Beliaev damping) of an excitation $E_q$ is impossible when $q>q^\ast$. 
The same phenomenon has been discussed for phonons in superfluid $^4\mathrm{He}$ \cite{anomalous1,anomalous2,Pitaevskii&Levinson} 
and one finds that $ q^\ast\sim p_0$ \cite{TsuchiyaGriffin}.
This disappearance of Beliaev damping for $q>q^\ast$ at $T=0$ should be easy to confirm experimentally.

In conclusion, assuming that only the lowest energy excitation band is thermally occupied, we have shown that Landau damping of Bogoliubov excitations in uniform optical lattices is only possible if they exhibit anomalous dispersion.

%\begin{acknowledgments} 
We thank David Luxat and Mike Smith for many helpful discussions.
S.T. is supported by JSPS of Japan while 
A.G. is supported by NSERC of Canada.

%Note the equation numbers in this appendix, produced with the
%subequations environment:
%\begin{subequations}
%\begin{eqnarray}
%E&=&mc, \label{appa}
%\\
%E&=&mc^2, \label{appb}
%\\
%E&\agt& mc^3. \label{appc}
%\end{eqnarray}
%\end{subequations}
%They turn out to be Eqs.~(\ref{appa}), (\ref{appb}), and (\ref{appc}).
%\newpage %Just because of unusual number of tables stacked at end

\end{document}